\documentclass[preprints,article,accept,pdftex,moreauthors]{Definitions/mdpi} 
\firstpage{1} 
\pubvolume{1}
\issuenum{1}
\articlenumber{0}
\pubyear{2023}
\copyrightyear{2023}
\externaleditor{{Academic Editor: Dimitris M. Christodoulou} 
} 
\datereceived{4 December 2022} 
\daterevised{14 February 2023} 
\dateaccepted{20 February 2023} 
\datepublished{ } 
\hreflink{https://doi.org/} 

\newcommand{\bhac}{\texttt{BHAC}}
\newcommand{\hamr}{\texttt{H-AMR}}
\newcommand{\bhoss}{\texttt{BHOSS}}
\newcommand{\ipole}{\texttt{IPOLE}}
\newcommand{\koral}{\texttt{KORAL}}
\newcommand{\gdet}{\sqrt{-g}}
\newcommand{\MBH}{M87$^*$}
\newcommand{\sgra}{Sgr~A$^*$}

\Title{Accretion Flow Morphology in Numerical Simulations of Black Holes from the ngEHT Model Library: The Impact of Radiation Physics}

\TitleCitation{Accretion Flow Morphology in Numerical Simulations of Black Holes from the ngEHT Model Library: The Impact of Radiation Physics}

\Author{\textls[-15]{Koushik Chatterjee $^{1,2,}$*\orcidA{}, Andrew Chael $^{3}$, Paul Tiede $^{1,2}$, Yosuke Mizuno $^{4,5,6}$, Razieh Emami $^{2}$, Christian Fromm $^{6,7,8}$,} Angelo Ricarte $^{1,2}$, Lindy Blackburn $^{1,2}$, Freek Roelofs $^{1,2}$, Michael D. Johnson $^{1,2}$, Sheperd S. Doeleman $^{1,2}$, \linebreak Philipp Arras $^{9,10}$, Antonio Fuentes $^{11}$, Jakob Knollm\"uller $^{10,12}$, Nikita Kosogorov $^{13,14}$, Greg Lindahl $^{2}$, \linebreak Hendrik M\"uller $^{8}$, Nimesh Patel $^{2}$, Alexander Raymond $^{1,2}$, Efthalia Traianou $^{11}$ and Justin Vega $^{2,15}$}

\AuthorNames{Koushik Chatterjee, Andrew Chael, Paul Tiede, Yosuke Mizuno, Razieh Emami, Christian Fromm, Angelo Ricarte, Lindy Blackburn, Freek Roelofs, Michael D. Johnson, Sheperd S. Doeleman, Philipp Arras, Antonio Fuentes, Jakob Knollm\"uller, Nikita Kosogorov, Greg Lindahl, Hendrik M\"uller, Nimesh Patel, Alexander Raymond, Efthalia Traianou and Justin Vega}

\AuthorCitation{Chatterjee, K.; Chael, A.; Tiede, P.; Mizuno, Y.; Emami, R.; Fromm, C.; Ricarte, A.; Blackburn, L.; Roelofs, F.; Johnson, M.D.; et al.}

\address{%
$^{1}$ \quad Black Hole Initiative, Harvard University, 20 Garden Street, Cambridge, MA 02138, USA\\
$^{2}$ \quad Center for Astrophysics, Harvard \& Smithsonian, 60 Garden Street, Cambridge, MA 02138, USA\\
$^{3}$ \quad Princeton Gravity Initiative, Princeton University, Jadwin Hall, Princeton, NJ 08544, USA\\
$^{4}$ \quad Tsung-Dao Lee Institute, Shanghai Jiao-Tong University, 520 Shengrong Road, Shanghai 201210, China\\
$^{5}$ \quad School of Physics \& Astronomy, Shanghai Jiao-Tong University, 800 Dongchuan Road, Shanghai 200240, China\\
$^{6}$ \quad Institut f\"ur Theoretische Physik, Goethe-Universit\"at Frankfurt, Max-von-Laue-Stra\ss e 1, D-60438 Frankfurt am Main, Germany\\
$^{7}$ \quad Institut f\"ur Theoretische Physik und Astrophysik, Universit\"at W\"urzburg, Emil-Fischer-Str. 31, D-97074 W\"urzburg, Germany\\
$^{8}$ \quad Max--Planck--Institut f\"ur Radioastronomie, Auf dem H\"ugel 69, D-53121 Bonn, Germany\\
$^{9}$ \quad Technical University Munich (TUM), Boltzmannstr. 3, 85748 Garching, Germany\\
$^{10}$\quad Max--Planck Institute for Astrophysics, Karl-Schwarzschild-Str. 1, 85748 Garching, Germany\\
$^{11}$\quad Instituto de Astrof\'isica de Andaluc\'ia-CSIC, Glorieta de la Astronom\'ia s/n, E-18008 Granada, Spain\\
$^{12}$\quad Excellence Cluster ORIGINS, Boltzmannstr. 2, 85748 Garching, Germany\\
$^{13}$\quad Moscow Institute of Physics and Technology, Institutsky per. 9, Dolgoprudny 141700, Russia\\
$^{14}$\quad Lebedev Physical Institute of the Russian Academy of Sciences, Leninsky Prospekt 53, 119991 Moscow, Russia\\
$^{15}$\quad Department of Physics, Northeastern University, 360 Huntington Ave, Boston, MA 02115, USA\\
}

\corres{Correspondence: koushik.chatterjee@cfa.harvard.edu (K.C.); achael@princeton.edu (A.C.); paul.tiede@cfa.harvard.edu (P.T.); mizuno@sjtu.edu.cn (Y.M.); razieh.emami\_meibody@cfa.harvard.edu (R.E.); christian.fromm@uni-wuerzburg.de (C.F.); angelo.ricarte@cfa.harvard.edu (A.R.); lblackburn@cfa.harvard.edu (L.B.); freek.roelofs@cfa.harvard.edu (F.R.); mjohnson@cfa.harvard.edu (M.D.J.); sdoeleman@cfa.harvard.edu (S.S.D.); parras@mpa-garching.mpg.de (P.A.); afuentes@iaa.es (A.F.); jakob@mpa-garching.mpg.de (J.K.); nakosogorov@gmail.com (N.K.); lindahl@pbm.com (G.L.); hmueller@mpifr-bonn.mpg.de (H.M.); npatel@cfa.harvard.edu (N.P.); awray314@gmail.com (A.R.); traianou@iaa.es (E.T.); justin.vega@cfa.harvard.edu (J.V.)}

\abstract{In the past few years, the Event Horizon Telescope (EHT) has provided the first-ever event horizon-scale images of the supermassive black holes (BHs) \MBH{} and Sagittarius A$^*$ (\sgra{}). The next-generation EHT project is an extension of the EHT array that promises larger angular resolution and higher sensitivity to the dim, extended flux around the central ring-like structure, possibly connecting the accretion flow and the jet. The ngEHT Analysis Challenges aim to understand the science extractability from synthetic images and movies to inform the ngEHT array design and analysis algorithm development. In this work, we compare the accretion flow structure and dynamics in numerical fluid simulations that specifically target \MBH{} and \sgra{}, and were used to construct the source models in the challenge set. We consider (1) a steady-state axisymmetric radiatively inefficient accretion flow model with a time-dependent shearing hotspot, (2) two time-dependent single fluid general relativistic magnetohydrodynamic (GRMHD) simulations from the \hamr{} code, \mbox{(3) a two-temperature} GRMHD simulation from the \bhac{} code, and (4) a two-temperature radiative GRMHD simulation from the \koral{} code. We find that the different models exhibit remarkably similar temporal and spatial properties, except for the electron temperature, since radiative losses substantially cool down electrons near the BH and the jet sheath, signaling the importance of radiative cooling even for slowly accreting BHs such as \MBH{}. We restrict ourselves to standard torus accretion flows, and leave larger explorations of alternate accretion models to future work.}


\begin{document}


\section{Introduction}

With the advent of the Event Horizon Telescope ([EHT;~\cite{EHT_M87_2019_PaperI, EHT_SgrA_2022_PaperI}), imaging the near-horizon structure of supermassive black holes (BHs) is now a reality. The primary targets of the EHT and the future next-generation EHT (or ngEHT\endnote{\url{https://www.ngeht.org/} (accessed on 22 February 2023)}) are \MBH{} (the supermassive BH in the elliptical galaxy M87;~\cite{EHT_M87_2019_PaperI}) and \sgra{} in the Galactic Center~\citep{EHT_SgrA_2022_PaperI}, which are two of the most well-studied low-luminosity active galactic nuclei. Extracting information about the event horizon-scale accretion flows in these two sources using the EHT's enormous resolving power is an active area of research. With the ngEHT, we will achieve unprecedented levels of angular resolution and sensitivity to low-flux regions, with the dynamic range in flux expected to increase to $\sim$1000 compared to the EHT's current dynamic range of $\sim$10 (e.g.,~\cite{Doeleman2019}). This would enable us to investigate the BH shadow shape with higher precision as well as provide a crucial connection between the accretion flow and the jet launching region. The expected advances in sensitivity require deeper investigations of feature extraction from simulated synthetic reconstructions of BH systems. Hence, we designed the ngEHT analysis challenges \endnote{\url{https://challenge.ngeht.org/} (accessed on 22 February 2023)}~\citep{Roelofs_ngEHT} to test our ability to capture the complex dynamics of gas and magnetic fields around \MBH{} and \sgra{} using the ngEHT reference array (e.g.,~\cite{Raymond:2021}) with various analysis methods.

Black hole accretion and jet physics have been intensively studied over the past few decades (e.g.,~\cite{Shakura:73,ree82,Narayan:95a,quataert:2000:CDAF,narayan03,mckinney06, kom07, tch11, narayanSANE2012, tch16}). In the context of \MBH{} and \sgra{}, we expect the accretion flow to be highly sub-Eddington, radiatively inefficient, and geometrically thick, popularly known as radiatively inefficient accretion flows (RIAFs). This accretion flow solution has been used to successfully model the multiwavelength spectrum of \sgra{} (e.g.,~\cite{Yuan:03}). On the other hand, semi-analytical models of jets are preferred to explain the spectrum of \MBH{} (e.g.,~\cite{lucchini19:M87}). Thus, these two sources already provide a means to probe two different components of BH accretion, namely, the inner accretion flow structure and turbulence in \sgra{} and the prominent jet feature in \MBH{}. The first three EHT numerical simulation papers~\citep{EHT_M87_2019_PaperV,EHT_M87_2019_PaperVIII,EHT_SgrA_2022_PaperV} already provide us with important clues about the horizon-scale conditions of these BH systems based on numerical simulations: (1) these BHs probably have non-zero spin, (2) the accretion disk is expected to have colder electrons than the jet sheath, and (3) the observations favor the presence of dynamically important magnetic fields close to the BH. All of these results point us toward the magnetically arrested disk (MAD;~\cite{igu03,narayan03}) state, an accretion mode \textls[-5]{where the BH magnetosphere becomes over-saturated with magnetic flux and exhibits quasi-periodic explosions of vertical magnetic field bundles. MAD flows around spinning BHs also have powerful relativistic jets, where the jet power can exceed the input accretion power~\citep{tch11}, which is a definite signature of BH spin energy extraction via the~\citet{bz77}} process.

Building on the semi-analytical RIAF models, the time-dependent general relativistic magneto-hydrodynamic (GRMHD) simulations have become important tools for deciphering BH accretion physics in a variety of astrophysical systems (e.g.,~\cite{Gammie:03, mckinney06,fragile07, tch11,Chael_2019, Porth:19, Narayan2022}). Indeed, the EHT regularly uses the libraries of GRMHD simulations to model the observed horizon-scale BH images of \MBH{} and \sgra{} as well as larger-scale jet images (such as for Centaurus A;~\cite{Janssen:2021}) in order to constrain the time-variable plasma properties. In designing the ngEHT reference array, it is therefore crucial to use GRMHD simulations for understanding the attainability of specific science goals, such as resolving the photon ring and the disk-jet connection region as well as tracing out time-variable features via the ngEHT analysis challenges. 

In this work, we discuss the numerical fluid simulations that were used as source models for the ngEHT analysis challenges. In particular, our objective is to compare between the models that incorporate increasingly complicated levels of accretion and electron physics, focusing on \MBH{} and \sgra{}. Our model set consists of a time-dependent shearing hotspot stitched to a steady-state RIAF solution, two standard GRMHD simulations of MAD accretion flows, a GRMHD MAD simulation with electron heating via incorporating two-temperature physics, and a fully radiative, two-temperature, GRMHD MAD simulation. We describe the equations and setup of our numerical models in Section~\ref{sec:sims}, show our comparison results in Section~\ref{sec:results}, and, finally, conclude in Section~\ref{sec:conclusions}.

\section{Numerical Simulations}
\label{sec:sims}

In this section, we provide a brief description of the semi-analytical stationary RIAF and shearing hotspot model as well as the (two-temperature/radiative) GRMHD simulations used for the ngEHT analysis challenges. 

\subsection{RIAF + Hotspot Solutions}
\label{sec:RIAF}
RIAF models attempt to describe the time and azimuthally averaged appearance of accretion flows. This is performed using a set of building blocks. The specific RIAF models used in the challenges are based on~\citet{Yuan:03, broderick_2006, Broderick_2010} approach. We decompose the accretion flow into a set of phenomenological models that describe the electron density, temperature, magnetic field, and velocity profile. We have a cylindrical coordinate system $x^\mu\equiv (R_{\rm cyl}, \varphi, z)$. The electron density profile is defined in terms of the cylindrical radius $R_{\rm cyl} = r|\sin(\theta)|$ and vertical displacement $z = r\cos(\theta)$, and is provided by 
\begin{equation}
    n_{e,\rm{X}}(R_{\rm cyl}, z) = n_{e,\rm X, 0} R_{\rm cyl}^{p_X}\exp\left(-\frac{z^2}{2h^2R_{\rm cyl}^2}\right)
\end{equation}
where $X$ denotes the population of electrons. The disk height $h$ is set to unity for this work. For the challenge dataset, we included both thermal synchrotron emitting ($X\equiv{\rm th})$, and non-thermal synchrotron ($X\equiv{\rm nth}$) emitting electrons. The thermal electrons have $n_{e,{\rm th}, 0}=1.3 \times 10^8 $ and $p_{\rm th} = -1.1$, while the non-thermal electrons are provided by $n_{e, {\rm nth}, 0} =  1.3\times 10^5$, $p_{\rm nth} =-2.02$. These numbers are from~\citet{Tiede2020} and are set to match the best-fit parameters for the \sgra{} spectrum from~\citet{Broderick:2011} and~\citet{Broderick:2016}.

The temperature profile of the thermal electrons is also provided by a radial power law with a Gaussian envelope describing the height:
\begin{equation}\label{eq:Te_riaf}
    T_{e}(t,r,\theta, \varphi) = T_{e,0}R_{\rm cyl}^{-0.84}\exp\left(-\frac{z^2}{2h^2R_{\rm cyl}^2}\right),
\end{equation}
where, for the challenge data, we set $T_{e,0} = 6.3\times 10^{10}$~K. 

Following~\citet{Pu2016}, we define the gas pressure by assuming that the ultra-relavistic protons are in roughly virial equilibrium with the gravitational force providing:
\begin{equation}\label{eq:riaf_gas}
    p_{\rm gas}(x^\mu) = \frac{m_p n_{e,th}(x^\mu)}{6}\frac{GM_{\rm BH}}{r},
\end{equation}
\noindent where $m_p$, $G$, and $M_{\rm BH}$ are the proton mass, gravitational constant, and the black hole mass.

For the local magnetic field, we then assume a constant $\beta = p_{\rm gas}/p_{\rm mag}=10$ plasma, which, combined with Equation~(\ref{eq:riaf_gas}), provide us the magnetic field strength. The orientation of the magnetic field is then provided by a purely toroidal configuration, relative to the plasma observer. Finally, we extract the emission and absorption coefficients from the synchrotron self-absorption model in~\citet{broderickblandford04} for both the thermal and non-thermal synchrotron electrons. For the non-thermal electrons, we use a power law prescription with radiation coefficients from~\citet{jonesodell}, and a photon spectral power law index of 1.25 (see~\cite{Tiede2020} for more details).

We follow the velocity field prescription from~\citet{Pu2016} to describe the accretion flow dynamics. Using the notation from~\citet{Tiede2020}, the velocity field $u^\mu = u^t v^\mu$ is provided by
\begin{equation}\label{eq:hotspot_velo}
\begin{aligned}
    v^r &= v^r_K + \alpha_{\rm vel}(v^r_{ff} - v^r_{K}) \\
    v^\theta &= 0 \\
    v^\varphi &= v^\varphi_K + (1-\kappa_{\rm vel})(v^\varphi_{ff} - v^\varphi_K),
\end{aligned}
\end{equation}
where $v^\mu_K$ denotes the Keplerian velocity field, $v^\mu_{ff}$ is the free-fall velocity field, and \linebreak $\alpha_{\rm vel}=0.01$ and $\kappa_{\rm vel}=1.0$, i.e., the hotspot rotates with Keplerian angular velocity and is weakly free-falling. For $\alpha_{\rm vel}=0, \kappa_{\rm vel}=1$, we have a Keplerian orbit, whereas, for $\alpha_{\rm vel}=1, \kappa_{\rm vel}=0$, we have free-fall. The remaining component $u^t$ is provided by the normalization condition $u^\mu u_\mu = -1$. The radial component outside the inner stable circular orbit (ISCO) is $v^r_K = 0$ as the disk is in steady-state. However, inside the ISCO, we use plunging geodesics that are specified by matching the angular momentum and energy at the ISCO. 

The hotspot evolution follows the model from~\citet{Tiede2020} (also see~\cite{broderick_2006}), where we assume that the hotspot travels passively along the accretion flow velocity field Equation~(\ref{eq:hotspot_velo}). This implies that the equation of motion is provided by the conservation of particle number Equation~(\ref{eq:continuity}). For the emission, we assume a non-thermal synchrotron hotspot, with an initial Gaussian density profile
\begin{equation}
    n_{e}(x^\mu) = n_0e^{-\Delta r^\mu r_\mu +(\Delta r_\mu v^\mu)^2/2R_s^2}, 
\end{equation}
where we have set $n_0 = 6\times 10^6$, hotspot size $R_s = 0.5\,r_{\rm g} = 0.5 GM_{\rm BH}/c^2$, and $\Delta r^\mu$ is the displacement from the hotspot center. 

\subsection{GRMHD Simulations}
\label{sec:GRMHD}

Over the previous two decades, multiple GRMHD codes have been developed and utilized to model black hole accretion and jet launching physics over long dynamical timescales. The wide usage of GRMHD simulations is particularly encouraging since this allows for verification of code-specific numerical choices that users usually have to make even while solving the same base set of GRMHD equations. Indeed, recently, there was a community-wide effort to benchmark these codes against each other for a standard problem: evolving a weakly magnetized torus of gas around a spinning black hole~\citep{Porth:19}. It was found that these codes largely provide similar solutions, though some disk quantities remain unconverged with increasing grid resolutions, suggesting more investigation is required. For this work, we employ three different GRMHD codes to probe black hole accretion, increasing the complexity of the equations solved at each step: (1) single fluid GRMHD simulations from the \hamr{} code~\citep{liska_hamr2020}, (2) a two-temperature single fluid GRMHD simulation from the \bhac{} code~\citep{porth17}, and (3) a two-temperature radiative GRMHD simulation from the \koral{} code~\citep{Sadowski:13_koral}.

First, we describe the set of GRMHD Equations (e.g., from~\cite{Gammie:03, Porth:19}). We have the conservation of particle number and energy-momentum:
\begin{eqnarray}
      \label{eq:continuity}
      \partial_{t}(\gdet \rho u^{t}) &=& -\partial_{i}(\gdet \rho u^{i}) \\
      \partial_{t}(\gdet T^t_{\nu}) &=& -\partial_{i}(\gdet T^i_{\nu}) +\gdet T^{\alpha}_{\beta}\Gamma^{\beta}_{\nu \alpha}.
\end{eqnarray}

Here, $\rho$ is the rest-mass gas density and can also be written in the form of $\rho=m n$, where $m$ is the mean rest--mass per particle and $n$ is the particle number density. We also have the four-velocity $u^{\mu}$, stress--energy tensor $T^{\mu}_{\nu}$, metric determinant $g\equiv det(g_{\mu\nu})$, and the metric connection $\Gamma^{\beta}_{\nu \alpha}$. Note that the index $t$ refers to the temporal component of the vector or tensor and $i$ denotes the spatial indices. 

The stress-energy tensor $T^{\mu}_{\nu}$ is provided as
\begin{equation}
    T^{\mu}_{\nu}=(\rho + U_{\rm gas} + p_{\rm gas} + 2p_{\rm mag})u^{\mu}u_{\nu} + (p_{\rm gas} + p_{\rm mag})\delta^{\mu}_{\nu} - b^{\mu}b_{\nu},
\end{equation}
\noindent where $U_{\rm gas}$ and $p_{\rm gas}$ are the gas internal energy and pressure, related by the ideal gas equation: $p_{\rm gas}=(\Gamma_{\rm gas}-1)U_{\rm gas}$ assuming a gas adiabatic index $\Gamma_{\rm gas}$. We also have the magnetic pressure $p_{\rm mag}=b^2/2$ and the magnetic field 4-vector $b^{\mu}$, which can be defined in terms of the magnetic field 3-vector $B^i$:
\begin{eqnarray}
      b^t &=& B^i u^{\mu} g_{i\mu} \\
      b^i &=& (B^i + b^t u^i)/u^t.
\end{eqnarray}

Here, we included a factor of $\sqrt{4\pi}$ into the definition of $B^i$. We evolve the magnetic field $B^i$ using the spatial components of the induction equation,
\begin{equation}
    \partial_{t}(\gdet B^i) = -\partial_{j}(\gdet (b^j u^i - b^i u^j)),
\end{equation}
\noindent while the temporal component provides the no monopoles constraint,
\begin{equation}
    \frac{1}{\gdet}\partial_i(\gdet B^i) = 0.
\end{equation}

These equations are numerically integrated in the conservative form~\citep{Gammie:03} to obtain the physically relevant quantities $\rho$, $U_{\rm gas}$, $u^{\mu}$, and $B^i$. We refer the reader to the corresponding code papers for more information on the numerical techniques used to evolve these equations over space and time.

In this work, we use two GRMHD simulations performed with the \hamr{} code, one targeting \MBH{} and the other \sgra{}. These simulations employ logarithmic Kerr--Schild coordinates and the grid resolutions are $N_r\times N_\theta \times N_\varphi = 580 \times 288 \times 512$ for the \MBH{} simulation and $348 \times 192 \times 192$ for the \sgra{} simulation. All the simulations in this work adopt the geometrical unit convention, $G=c=1$ and using $M_{\rm BH}=1$, normalizing the length scale to the gravitational radius $r_{\rm g}=GM_{\rm BH}/c^2$. The \MBH{} GRMHD simulation evolves a MAD flow around a black hole with spin $a=0.9375$. The \sgra{} model also simulates a MAD flow but around a black hole with spin $a=1/2$. \hamr{} uses outflowing radial boundary conditions (BCs), transmissive polar BCs, and periodic azimuthal BCs (for more details, see~\cite{liska_tilt_2018}). 

Since GRMHD simulations are scale-free, we determine the gas density code-to-CGS units conversion factor (hereafter, ``density scaling'') by raytracing the simulation at 230~GHz for a target source and flux. We use the general relativistic ray-tracing codes \bhoss{}~\citep{Younsi:19_polarizedbhoss} and \ipole{}~\citep{Moscibrodzka:2018} to compute images at 230 GHz and set the compact flux to be approximately 0.5 Jy for \MBH{}~\citep{EHT_M87_2019_PaperI} and 2.4 Jy for \sgra{}~\citep{EHT_SgrA_2022_PaperI}. We use the black hole masses and distances of $M_{\rm BH} =6.2\times10^9 M_{\odot}$ and $D_{\rm BH}=16.9$~Mpc for \MBH{} (e.g.,~\cite{Gebhardt:2011,EHT_M87_2019_PaperV} and references therein) and $M_{\rm BH}=4.14\times10^6 M_{\odot}$ and $D_{\rm BH}=8.127$~kpc for \sgra{}~\citep{Do_2019,Gravity:2019,EHT_SgrA_2022_PaperV}. 

GRMHD simulations evolve a single temperature fluid. At the low accretion rates seen in \MBH{} and \sgra{}, Coulomb coupling between ions and electrons is inefficient and, therefore, the two particle species are not in thermal equilibrium. Since ions are much heavier than electrons, the ions dominate the single fluid thermodynamics evolved in GRMHD simulations. Hence, to calculate the radiative output from GRMHD simulations, we calculate the electron temperature $T_{\rm e}$ using sub-grid models such as the $R-\beta$ prescription~\citep{Moscibrodzka_2016} based on local gas plasma-$\beta$ ($\equiv p_{\rm gas}/p_{\rm mag}$):

\begin{eqnarray}
    T_{\rm e} &=& \frac{2 m_{\rm p}U_{\rm gas}}{3k_{\rm B}\rho (2 + R)}, \\
     {\rm where,\,\,} R &=& \frac{1+R_{\rm high} \beta^2}{1+\beta^2}.
     \label{eq:Rbeta}
\end{eqnarray}

For the ngEHT analysis challenges, we select $R_{\rm high}$ values of 160 and 40 and source inclinations of $163^{\circ}$ and $50^{\circ}$ for the \MBH{} and \sgra{} simulations, respectively. We assume a thermal relativistic Maxwell--J\"uttner distribution for describing the electron energy distribution in the \sgra{} model, and a hybrid thermal$+$non-thermal $\kappa-$distribution (e.g.,~\cite{Xiao:2006, Davelaar:18}) for the \MBH{} model. The model images are shown in~\citet{Roelofs_ngEHT}.

\subsection{Two-Temperature Physics}
\label{sec:two-temp}

Two-temperature GRMHD (or 2t-GRMHD) simulations (e.g.,~\cite{ressler_2015, Dexter_2020}) evolve the ion and electron entropy equation separately and, hence, provide the ion and electron temperature in a self-consistent manner. The main advantage of this method is that we remove the electron temperature as a free parameter when constructing images. However, we do have to make a choice about the sub-grid prescription that determines the fraction of local dissipative energy that heats the electrons. There are two heating mechanisms that are thought to be applicable to global accretion flows: turbulent heating~\citep{Howes:2010, Kawazura:2019}, and magnetic reconnection~\citep{Werner:2018, Rowan:2017}. For the ngEHT analysis challenges, we focus only on one simulation with reconnection heating, from~\citet{Mizuno_2021}, as heating via magnetic reconnection in equatorial current sheets formed in magnetically arrested flows, captured by the~\citet{Rowan:2017} heating model, is arguably more important than heating via small-scale turbulent eddies that are more prevalent in weakly magnetized disks.    

We assume that the number densities and velocities of ions and electrons are equal, i.e., $n_{\rm i}=n_{\rm e}=n$ and $u^{\mu}_{\rm i}=u^{\mu}_{\rm e}=u^{\mu}$, thus maintaining charge neutrality. The electron entropy equation is provided as

\begin{equation}
    \partial_{\mu} (\gdet \rho u^{\mu} s_{\rm e}) = \frac{\gdet (\Gamma_{\rm e} - 1)}{\rho^{\Gamma_{\rm e} -1}}f_{\rm e}Q,
\end{equation}

\noindent where the electron entropy is $s_{\rm e} = p_{\rm e}/\rho^{\Gamma_{\rm e}}$, with $p_{\rm e}$ and $\Gamma_{\rm e}$ as the electron pressure and adiabatic index. The total heating rate $Q$ is calculated by comparing the total internal energy of the gas and the internal energy obtained from the electron entropy conservation Equation (see~\cite{ressler_2015} for more details). The fraction of dissipative heating that contributes to electron heating is provided by $f_{\rm e}$. For this particular simulation, $f_{\rm e}$ is designed to capture electron/ion heating via magnetic reconnection from~\citet{Rowan:2017}:

\begin{equation}
    f_{\rm e} = \frac{1}{2} \exp \Big[ -\frac{1-\beta/\beta_{\rm max}}{0.8+\sigma_{\rm h}^{0.5}} \Big],
\end{equation}

\noindent where $\beta_{\rm max} = 1/4\sigma_{\rm h}$, defined using the hot gas magnetization $\sigma_{\rm h} = b^2/\rho E_{\rm th}$ and the specific gas enthalpy $E_{\rm th}=1+\Gamma_{\rm gas} p_{\rm g}/[\rho(\Gamma_{\rm gas}-1)]$.

The 2t-GRMHD simulation from~\citet{Mizuno_2021} assumes modified Kerr--Schild coordinates and a black hole spin of $0.9375$. The grid resolution is $384 \times 192 \times 192$. The accretion mode is of a magnetically arrested flow and the simulation is raytraced (using \bhoss{}) once the near-horizon flow has reached steady state. The target source is \MBH{}, assuming a black hole mass of $M_{\rm BH}=6.5\times10^9 M_{\odot}$ and distance of 16.9 Mpc~\citep{EHT_M87_2019_PaperI}. The accretion rate is normalized such that the 230 GHz compact flux density is 0.8 Jy. We assume a thermal electron distribution everywhere except in the jet sheath where we adopt a $\kappa-$distribution. More details about the image are provided in~\citet{Roelofs_ngEHT}.

\subsection{Radiative GRMHD}
\label{sec:GRRMHD}

Two temperature GRMHD simulations do not include radiative cooling and, hence, are thought to be appropriate for low luminosity supermassive black holes such as \MBH{} and \sgra{}. To verify this assumption, we consider a two-temperature radiative GRMHD (2t-GRRMHD hereafter) simulations from~\citet{Chael_2019}. This simulation accounts for self-consistent radiation physics, incorporating both particle heating via magnetic reconnection (as in Section~\ref{sec:two-temp}) and radiative cooling via bremsstrahlung, synchrotron, Compton, and Coulomb losses. The simulation is run using the 2t-GRRMHD code \koral{}~\citep{Sadowski:13_koral, Sadowski:15_photon,Chael_2018}, which evolves a two-temperature magnetized fluid and treats the radiation field as a second fluid~\citep{Sadowski:2017}. The conservation equations solved in 2t-GRRMHD are different from that of~GRMHD:
\begin{equation}
    (T^{\mu}_{\nu} + R^{\mu}_{\nu})_{;\,\,\, \mu} =0,
\end{equation}
\noindent where $R^{\mu}_{\nu}$ is the frequency-integrated radiation field, defined as,
\begin{equation}
    R^{\mu}_{\nu} = \frac{4}{3} \overline{E} u^{\mu}_{\rm R} u_{\nu \rm R} + \frac{1}{3} \overline{E}\delta^{\mu}_{\nu}.
\end{equation}

Here, the radiation field is described by its rest-frame energy density $\overline{E}$ and four-velocity $u^{\mu}_{\rm R}$ following the M1 closure scheme. The ion and electron entropy equations~are
\begin{eqnarray}
      T_{\rm e} (ns_{\rm e}u^{\mu})_{;\mu} &=& f_{\rm e}q^{\rm v} + q^{\rm C} - G, \\
      T_{\rm i} (ns_{\rm i}u^{\mu})_{;\mu} &=& (1-f_{\rm e})q^{\rm v} - q^{\rm C},
\end{eqnarray}
\noindent where $q^{\rm v}$ is the dissipative heating rate and $q^{\rm C}$ is the Coulomb coupling rate that captures the exchange of energy between ions and electrons. The heating fraction $f_{\rm e}$ is from~\citet{Rowan:2017} (see \citet{Chael_2018} for more details), same as the 2t-GRMHD simulation. Finally, $G$ is the radiative cooling rate~\citep{Sadowski:13_koral}. For further details about the equations, see~\citet{Sadowski:13_koral,  Sadowski:2017, Chael_2018}. 

The simulation assumes a black hole spin of $a=0.9375$ and mass $M_{\rm BH}=6.2\times10^9 M_{\odot}$, targeting \MBH{}. The gas density is scaled to physical CGS units such that the compact emission at 230 GHz is roughly 0.98 Jy~\citep{Doeleman2012,Akiyama:2015}. The simulation uses modified Kerr--Schild coordinates with a grid resolution of $N_r\times N_\theta \times N_\varphi = 288 \times 224 \times 128$. See~\citet{Chael_2019} for more details about the simulation, while not utilized for the ngEHT analysis challenges, we included this simulation in this work since this model captures the coupling between gas and radiation, necessary for black holes accreting close the Eddington limit. Further, this model has been used in previous ngEHT reference array papers~\citep{Doeleman2019, Raymond:2021}.

\section{Results}
\label{sec:results}

We perform a series of comparisons focused on the time-evolution of horizon-scale quantities and radial dependence of disk and jet properties. The diagnostics are chosen such that any trends we find can inform EHT/ngEHT science applications, such as horizon-scale morphology and variability of the accretion flow. Further, the quantities are similar to those reported in the GRMHD code comparison project~\citep{Porth:19} and, thus, can be directly compared. There is a total of five models: three (2t-radiative) GRMHD simulations targeting M87 *, and one RIAF solution and one GRMHD simulation for Sgr A *. We further note that all three numerical simulations of M87 * have the same BH spin, favoring direct comparisons of the horizon-scale gas properties.

\subsection{Temporal Behavior of Horizon Fluxes}
\label{sec:time}

\begin{figure}[H]
    \includegraphics[width=\columnwidth]{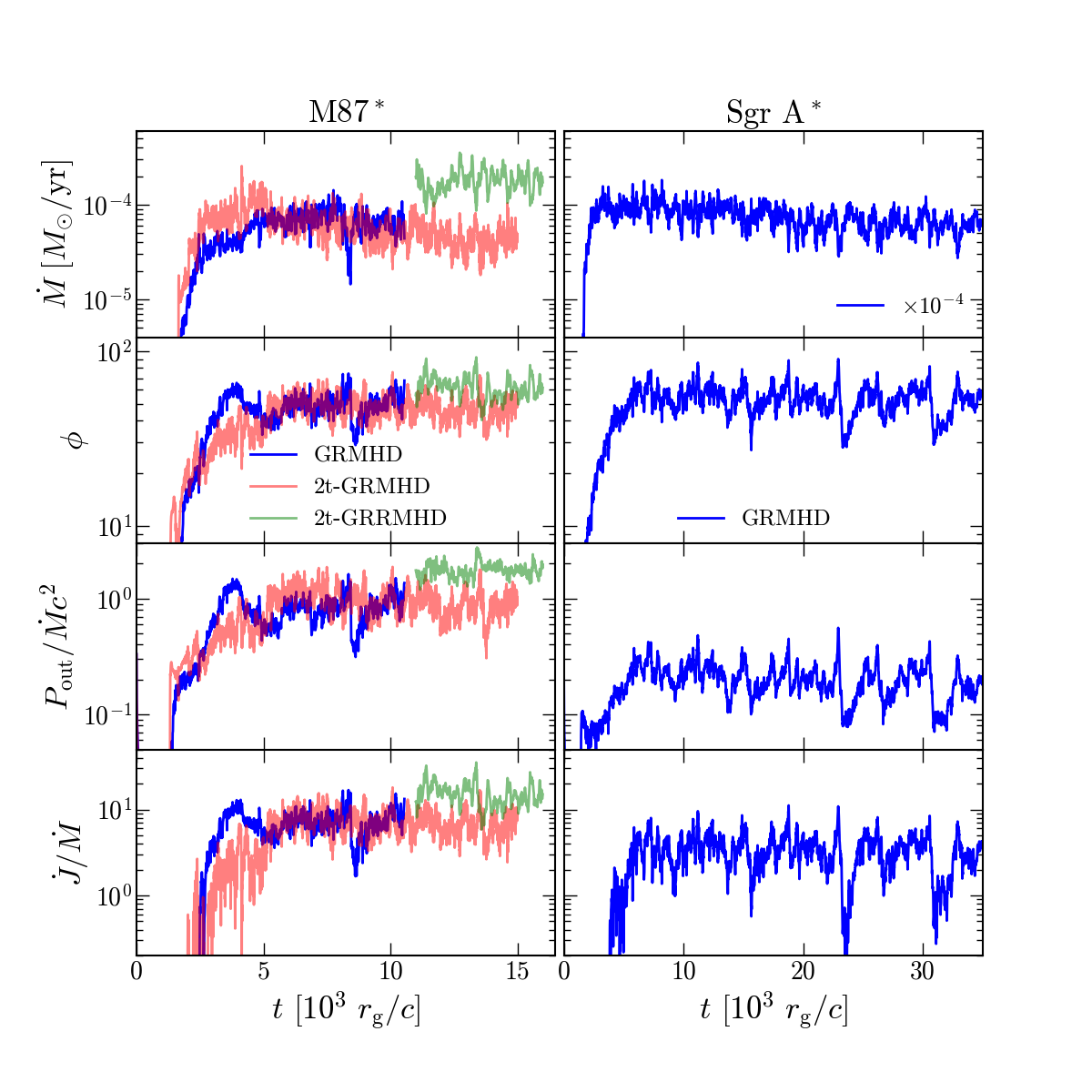}
    \vspace{-30pt}
	\caption{We show the mass accretion rate $\dot{M}$, dimensionless magnetic flux $\phi\equiv\Phi/\sqrt{\dot{M}}$, the outflow efficiency $P_{\rm out}/\dot{M}c^2=1-\dot{E}/\dot{M}c^2$, and specific radial flux of the angular momentum $\dot{J}/\dot{M}$ over time. Values are calculated at the event horizon.}
    \label{fig:Horizon_fluxes}
\end{figure}

We calculate the mass, magnetic, angular momentum and energy fluxes in the radial direction as follows:

\begin{eqnarray}
{\rm Mass:\,\,}& \dot{M} &=\int^{2\pi}_{0}\int^{\pi}_{0}\, (-\rho \, u^r) \, \sqrt{-g} \, d\theta \, d\varphi\,, \\
{\rm Magnetic:\,\,}& \Phi &=\frac{\sqrt{4\pi}}{2}\int^{2\pi}_{0}\int^{\pi}_{0}\, |B^r| \, \sqrt{-g} \, d\theta \, d\varphi\,, \\
{\rm Ang. Mom.:\,\,}& \dot{J} &=\int^{2\pi}_{0}\int^{\pi}_{0}\, T^r_{\varphi} \, \sqrt{-g} \, d\theta \, d\varphi\,, \\
{\rm Energy:\,\,}& \dot{E} &=\int^{2\pi}_{0}\int^{\pi}_{0}\, (T^r_t) \, \sqrt{-g} \, d\theta \, d\varphi\,, 
\end{eqnarray}

\noindent where all quantities are calculated at the event horizon radius $r_{\rm hor}=r_{\rm g} (1+\sqrt{1-a^2})$. We note that there could be a substantial contribution of density floors when calculating the mass accretion rate for MAD systems. However, this radius was chosen for simplicity when comparing to previous simulations in the literature. Figure~\ref{fig:Horizon_fluxes} shows the mass accretion rate $\dot{M}$ in units of solar masses per year ($M_{\odot}$/yr), the dimensionless magnetic flux $\phi=\Phi/\sqrt{\dot{M} r_{\rm g}^2 c }$, the outflow power $P_{\rm out} = \dot{M}c^2-\dot{E}$, and the specific angular momentum flux $\dot{J}/\dot{M}$ for simulations targeting M87 * and Sgr A *. The RIAF solution being a steady-state solution is excluded from this section (though the hotspot evolves with time). The quantities from the 2t-GRRMHD simulation are only shown for $(11-16)\times 10^3\,\,r_{\rm g}/c$, i.e., the time period over which the simulation was raytraced in~\citet{Chael_2019}. Remarkably, despite the difference in electron the physics complexity, the simulations behave very similarly. The factor of 2 difference in $\dot{M}$ between the \MBH{} non-radiative simulations and the 2t-GRRMHD simulation can be explained by the lower electron temperatures in the near-horizon accretion flow due to radiative cooling (see Section~\ref{sec:disk}) as well as the higher 230 GHz flux normalization used for the radiative model.

\begin{figure}[H]
    \includegraphics[height=2.12in, trim= 0mm 0mm 0mm 0mm, clip=true]{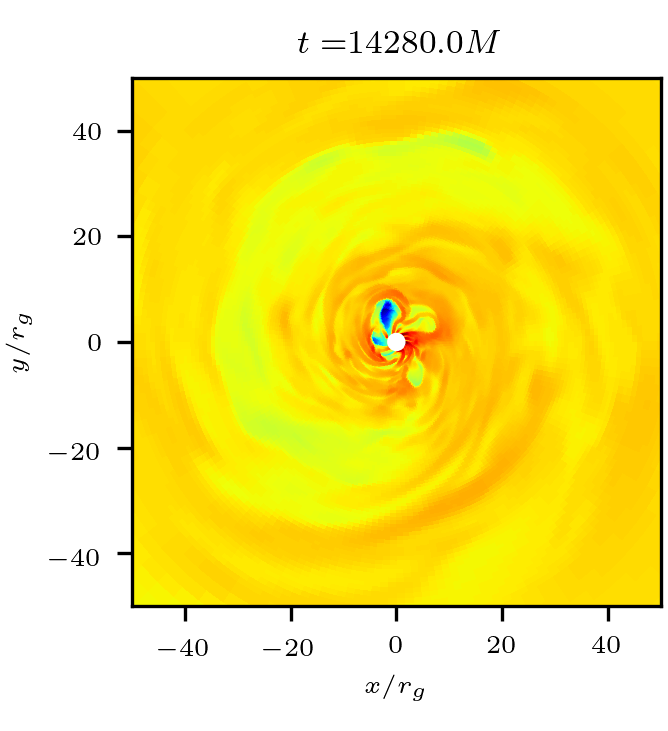}
    \includegraphics[height=2.12in]{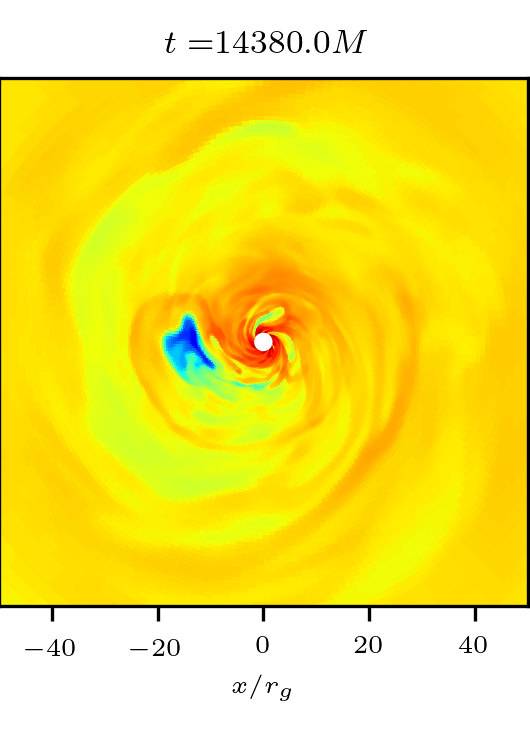}
    \includegraphics[height=2.12in]{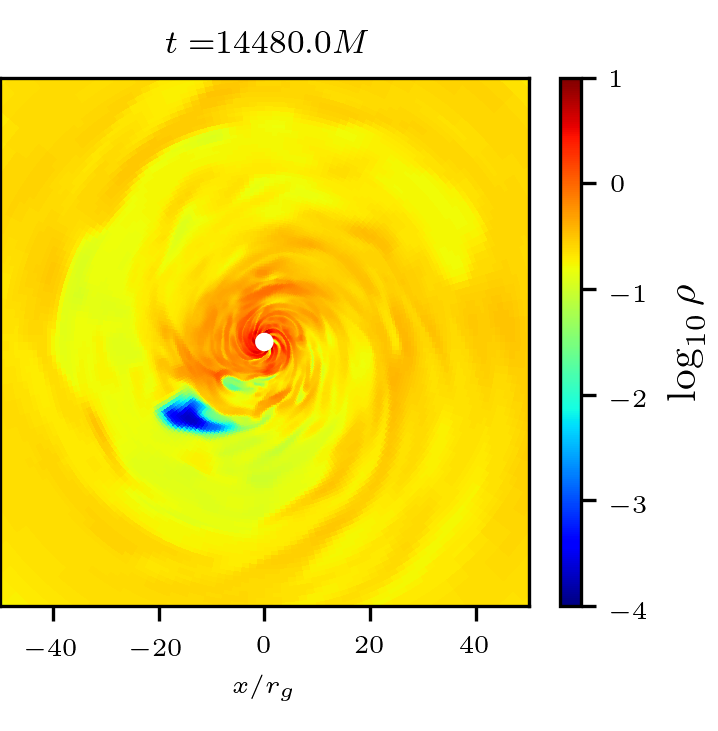}
    \includegraphics[height=2.12in]{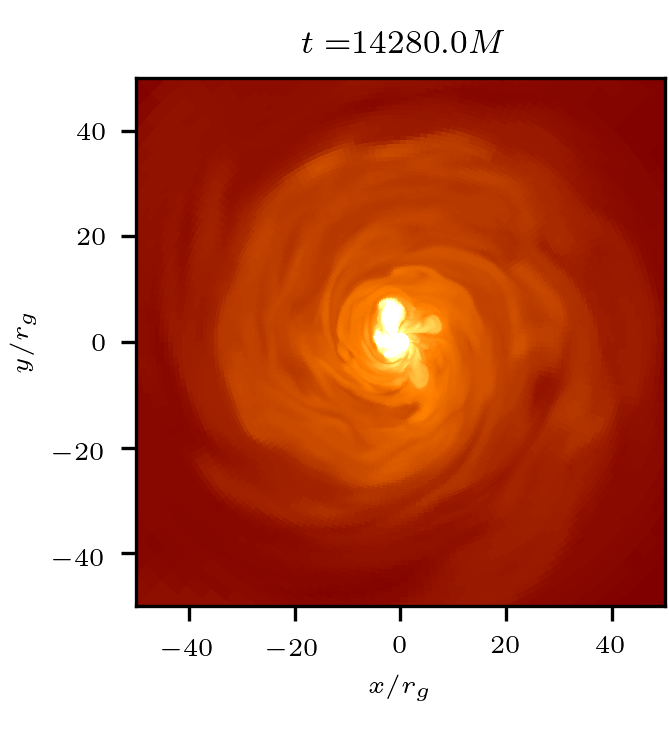}
    \includegraphics[height=2.12in]{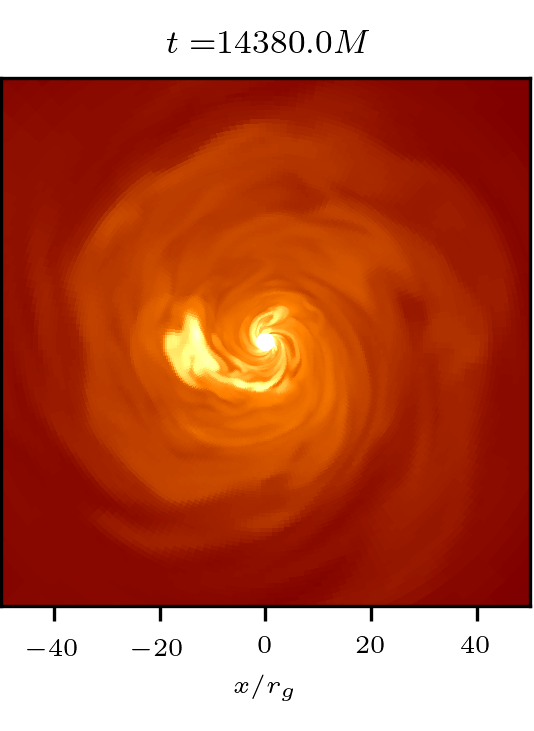}
    \includegraphics[height=2.12in]{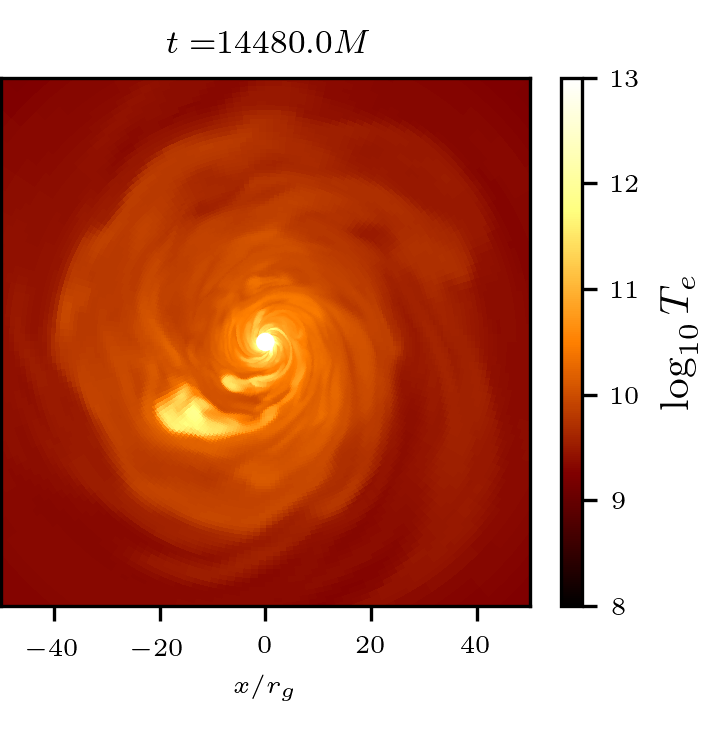}

	\caption{We show equatorial cross-sections of gas density $\rho$ (in arbitrary code units) and electron temperature $T_e$ (in Kelvin) during a magnetic flux eruption in the 2t-GRMHD simulation of \MBH{}. Magnetic flux eruptions, a characteristic feature of magnetically arrested disks, eject bundles of vertical magnetic fields filled with relativistically hot, low-density plasma. These flux bundles spiral around the black hole and may be responsible for high-energy flares~\citep{Ripperda2022}. The time unit $M$ is equivalent to $r_g/c$.}
    \label{fig:flux_eruption}
\end{figure}

The accretion rate in all simulations show large variation with quasi-periodic sharp drops. These drops in $\dot{M}$ occur due to the emergence of magnetic flux eruptions, a characteristic feature of the magnetically arrested disks~\citep{Porth:2020:NIR_MAD,Begelman2022, Ripperda2022, Chatterjee:2022}. These eruptions also lower the value of $\phi$ since magnetic flux bundles escape from the vicinity of the BH, carrying away the magnetic flux accumulated in the BH magnetosphere. We see that $\phi$ often crosses the magnetic flux saturation value of 50~\citep{tch11}, overwhelming the BH magnetosphere with strong magnetic fields that eventually reconnect and trigger flux eruptions (see~\cite{Ripperda2022} for the detailed mechanism). Figure~\ref{fig:flux_eruption} shows a series of equatorial snapshots from the \MBH{} 2t-GRMHD simulation where we follow a particular magnetic flux eruption by the rotating gas that is hot ($T_e\gtrsim10^{12}$~K), has low density ($\rho<-2$ in code units), and primarily consists of vertical field lines. As these field line bundles move out and interact with the disk, they (1) hinder accretion, lowering $\dot{M}$, (2) remove magnetic flux from near the BH, lowering the jet power, and (3) push gas outwards, reducing the inward angular momentum flux. Curiously, we see larger drops in the specific angular momentum flux for the \sgra{} GRMHD model. This is possibly due to the smaller BH spin ($a=0.5$ as opposed to $0.9375$ for the \MBH{} models) as the weakly powered jet does not carry away angular momentum as efficiently as the higher BH spin models and flux eruptions play a bigger role in regulating disk angular momentum transport. Additionally, the reconnection events that trigger these eruptions accelerate electrons to higher energies, and are, thus, crucial for understanding flare activity in BH sources. 

\vspace{-9pt} 

\begin{table}[H] 
\caption{GRMHD simulations considered in this work. Simulation grid resolution, simulation time period over which raytracing was performed, modulation index (MI) of the mass accretion rate $\dot{M}$, dimensionless magnetic flux $\phi$, outflow efficiency $P_{\rm out}/\dot{M}c^2$, and the specific angular momentum flux $\dot{J}/\dot{M}$ for each GRMHD model. The MI is calculated over the final $5000\,\,r_{\rm g}/c$ in runtime and at the event horizon (see Figure~\ref{fig:Horizon_fluxes}).}
\newcolumntype{C}{>{\centering\arraybackslash}X}
\begin{tabularx}{\textwidth}{lcccccc}
\toprule
\textbf{Model}	& \textbf{Grid Resolution}  & \textbf{Sim. Time} & \textbf{MI(}\boldmath{$\dot{M}$}\textbf{)}	& \textbf{MI(}\boldmath{$\phi$}\textbf{)} & \textbf{MI} & \textbf{MI}\\
\textbf{Name}	& ($N_{r}\times N_{\theta}\times N_{\varphi}$)  & \textbf{(}\boldmath{$\times 10^3\,\,r_{\rm g}/c$}\textbf{)} &	&  & \textbf{(}\boldmath{$P_{\rm out}/\dot{M}c^2$}\textbf{)} & \textbf{(}\boldmath{$\dot{J}/\dot{M}$}\textbf{)}\\
\midrule
\MBH{} GRMHD		& $580\times 288 \times 512$ & 5.6--10.6 & 0.27		    & 0.15   & 0.26   & 0.33\\
\MBH{} 2t-GRMHD		& $348\times 192 \times 192$ & 10--15    & 0.29			& 0.14   & 0.25   & 0.31\\
\MBH{} 2t-GRRMHD	& $288\times 224 \times 128$ & 11--16    & 0.28			& 0.14   & 0.14   & 0.31\\
\sgra{} GRMHD		& $348\times 192 \times 192$ & 30--35    & 0.23			& 0.21   & 0.39   & 0.57\\
\bottomrule
\end{tabularx}
\label{tab:MI}
\end{table}

To quantify the time-variability of the horizon-fluxes, we calculate the modulation index MI, which is defined as the ratio of the standard deviation and the mean of the quantity over time~\citep{EHT_SgrA_2022_PaperV}. We show the MI for the different fluxes in Table~\ref{tab:MI}. The MI($\dot{M}$) is usually a good proxy for the variability of the sub-millimeter (sub-mm) emission in these slowly accreting optically thin black hole sources (e.g.,~\cite{Chatterjee:2021}). The MI($\dot{M}$) values we see from the simulations are $\sim$0.23--0.29 and are larger than expected from \sgra{} 230 GHz lightcurves (where MI $\sim$ 0.1;~\cite{Wielgus:2022_SgrALC}). This suggests that careful analysis of the electron distribution function is needed to understand if we are substantially over-predicting the 230~GHz lightcurve variability. Further, in general, weakly magnetized accretion flows exhibit lower MI($\dot{M}$) values due to the absence of flux eruptions, which suggests that further study of the accretion mode in \sgra{} is also necessary. It is encouraging to note that our MI values for $\dot{M}$ and $\phi$ are consistent with the MI values from longer time-evolved GRMHD simulations of $a=0.9$ BHs in~\citet{Narayan2022}, indicating that our simulations are sufficiently converged with respect to horizon-scale quantities.

\subsection{Disk-Averaged Quantities}
\label{sec:disk}

Here, we calculate the disk-averaged properties of each model, namely gas density $\rho$, thermal pressure $p_{\rm gas}$, magnetic pressure $p_{\rm mag}$, radial velocity $|v_r|$, azimuthal velocity $|v_{\varphi}|$, angular momentum $u_{\varphi}$, disk scale height $h/r$, ion temperature $T_i$, and the electron temperature $T_e$. We define disk-averaging of a quantity $q$ as
\begin{equation}
    \langle q \rangle (r,t) = \frac{\int^{2\pi}_{0}\int^{\pi}_{0}\, \rho \, q \, \sqrt{-g} \, d\theta \, d\varphi}{\int^{2\pi}_{0}\int^{\pi}_{0}\, \rho \, \sqrt{-g} \, d\theta \, d\varphi},
\end{equation}
where $q \in \{ \rho, p_{\rm gas}, p_{\rm mag}, |v_r|, |v_{\varphi}| , u_{\varphi}, h/r, T_i [\rm Kelvin], T_e [\rm Kelvin]  \}$. Further definitions follow:
 
 \begin{eqnarray}
       p_{\rm gas} &=& (\Gamma_{\rm ad}-1)u, \\
       p_{\rm mag} &=& b^{\mu}b_{\mu}/2, \\
       |v_i| &=& \sqrt{v^i v^i g_{ii}}, \\
       {\rm where,\,\,} v^i &=& u^i/u^t, \nonumber \\
       h/r &=& |\theta - \pi/2|,
 \end{eqnarray}
where $\Gamma_{\rm ad}$ and $u$ are the adiabatic index and the internal energy of the gas. 

\vspace{-40pt}
\begin{figure}[H]
    \includegraphics[width=\columnwidth]{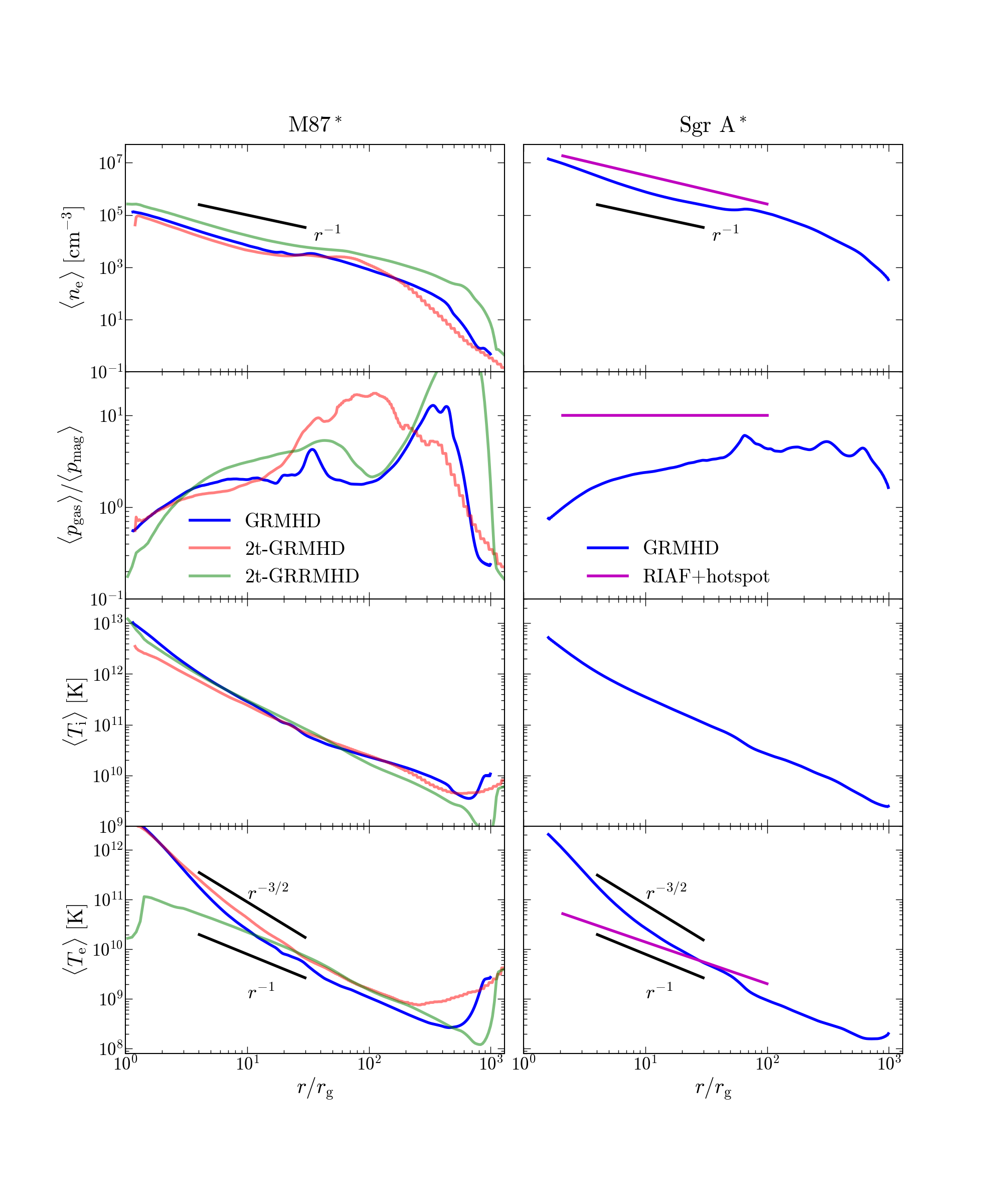}
    \vspace{-40pt}
	\caption{We show the radial profiles of gas density $\rho$, plasma$-\beta$, proton temperature $T_{\rm p}$, and electron temperature $T_{\rm e}$. Quantities are disk-averaged and time-averaged over the raytracing period.}
    \label{fig:disk_profiles_1}
\end{figure}

\begin{figure}[H]
    \includegraphics[width=\columnwidth]{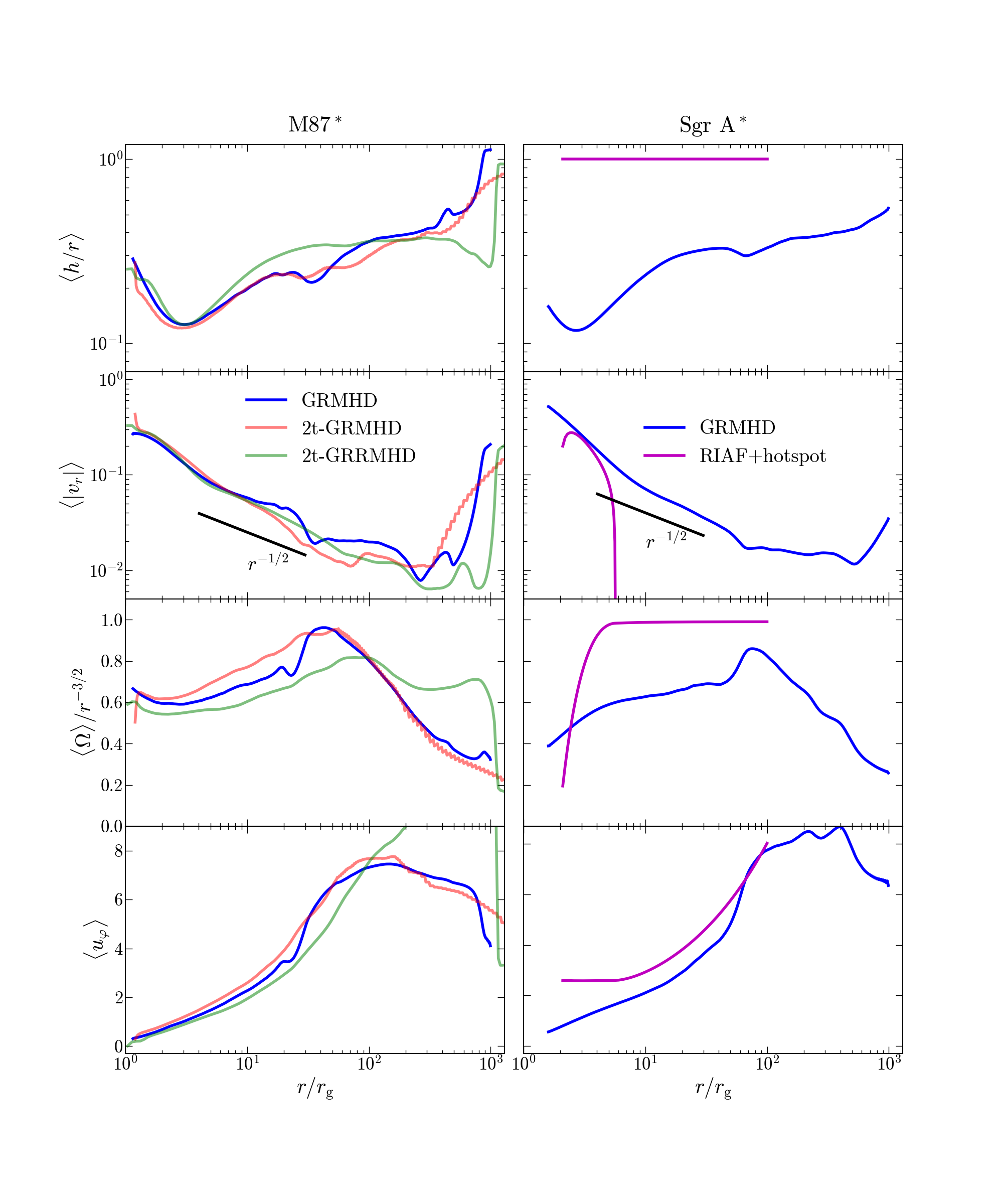}
    \vspace{-30pt}
	\caption{We show the radial profiles of disk scale height $h/r$, radial velocity $|v_r|$, angular velocity $\Omega$, and specific angular momentum $u_{\varphi}$. Quantities are disk-averaged and time-averaged over the raytracing period.}
    \label{fig:disk_profiles_2}
\end{figure}

 Figure~\ref{fig:disk_profiles_1} and \ref{fig:disk_profiles_2} show the respective disk-averaged radial profiles for each model, including the \sgra{} RIAF solution. The density profiles in the inner few tens of $r_{\rm g}$ converge roughly to a $n_{\rm e}\propto r^{-1}$ profile, matching the RIAF density profile as well as the longer time-evolved MAD simulations~\citep{Chatterjee:2022}. The \MBH{} 2t-GRRMHD density is larger by a factor of $\approx 2$ from the GRMHD/2t-GRMHD, as is expected from the difference in the mass accretion rate (Figure~\ref{fig:Horizon_fluxes}). The 2t-GRRMHD simulation exhibits a slightly more magnetized inflow within the inner $2\,\,r_{\rm g}$, but, overall, the GRMHD simulations have a similar plasma-$\beta\equiv p_{\rm gas}/p_{\rm mag}$ disk profile. The stronger magnetic field seen in the 2t-GRRMHD model could explain the higher values of the horizon magnetic flux seen in Figure~\ref{fig:Horizon_fluxes}. The RIAF model assumes a constant disk plasma-$\beta=10$, (see Section~\ref{sec:RIAF}), which is substantially higher when compared to the MAD GRMHD models. This value of plasma-$\beta$ is chosen in order to match the observed 230 GHz flux density of \sgra{}. As we see from the disk scale height in Figure~\ref{fig:disk_profiles_2}, the RIAF model has a much thicker disk than the GRMHD models, and, therefore, produces a lot more sub-mm emission even with a low electron temperature and weak magnetic field strength.

\vspace{-30pt}
 
 Next, we see that the disk-averaged electron temperature $T_{\rm e}$ in the 2t-GRRMHD \MBH{} model is more than an order of magnitude lower than the other GRMHD models within the inner $10\,\, r_{\rm g}$, but actually matches the \sgra{} RIAF $T_{\rm e}$ profile and has a shallower profile $T_{\rm e} \propto r^{-1}$ instead of $r^{-3/2}$. It is also interesting to note that the disk ion temperatures $T_{\rm i}$ are very similar in all the GRMHD simulations shown here. Therefore, despite the same reconnection-driven heating mechanism captured in both the 2t-GRMHD and the 2t-GRRMHD models, the radiative cooling of hot electrons plays a crucial role in determining the eventual $T_{\rm e}$. Due to the low $T_{\rm e}$, the required accretion rate normalization is higher in the 2t-GRRMHD, as we noted in the previous subsection. 

 In Figure~\ref{fig:disk_profiles_2}, we show the average disk scale height $h/r$, the radial and angular velocities ($v_r$ and $\Omega$ and the specific angular momentum $u_{\varphi}$). The MAD simulations all show very similar disk properties. The $\langle h/r\rangle \approx$ 0.1--0.3 with a sharp increase within $3\,\,r_{\rm g}$ where the inflow becomes vertically supported by strong poloidal magnetic fields. The radial velocity has a profile of $r^{-1/2}$, similar to the scaling relation found in ADAF solutions assuming a constant viscosity parameter $\alpha$~\citep{nar94,nar98}. The $\alpha$ parameter profile depends on how magnetized the accretion flow is, with $\alpha\propto r^{-1}$ for the weakly magnetized flows and close to constant for the MAD-like flows (e.g.,~\cite{liska_tor_2019,Chatterjee:2022}). We also see highly sub-Keplerian angular velocity profiles in the GRMHD models, typical for magnetically supported disks. For the RIAF model, the RIAF disk is not infalling and has a constant Keplerian angular velocity. Instead, the hotspot, added to the RIAF solution, undergoes shearing and disappears into the BH with a radial velocity similar to the values found in the GRMHD MAD disks. This occurs because the hotspot is designed to travel along plunging geodesics (see Section~\ref{sec:RIAF}), similar to rapid gas infall close to the BH in the GRMHD models. The angular momentum in the GRMHD models looks sub-Keplerian as expected for MADs.

\begin{figure}[H]
    \includegraphics[width=\columnwidth]{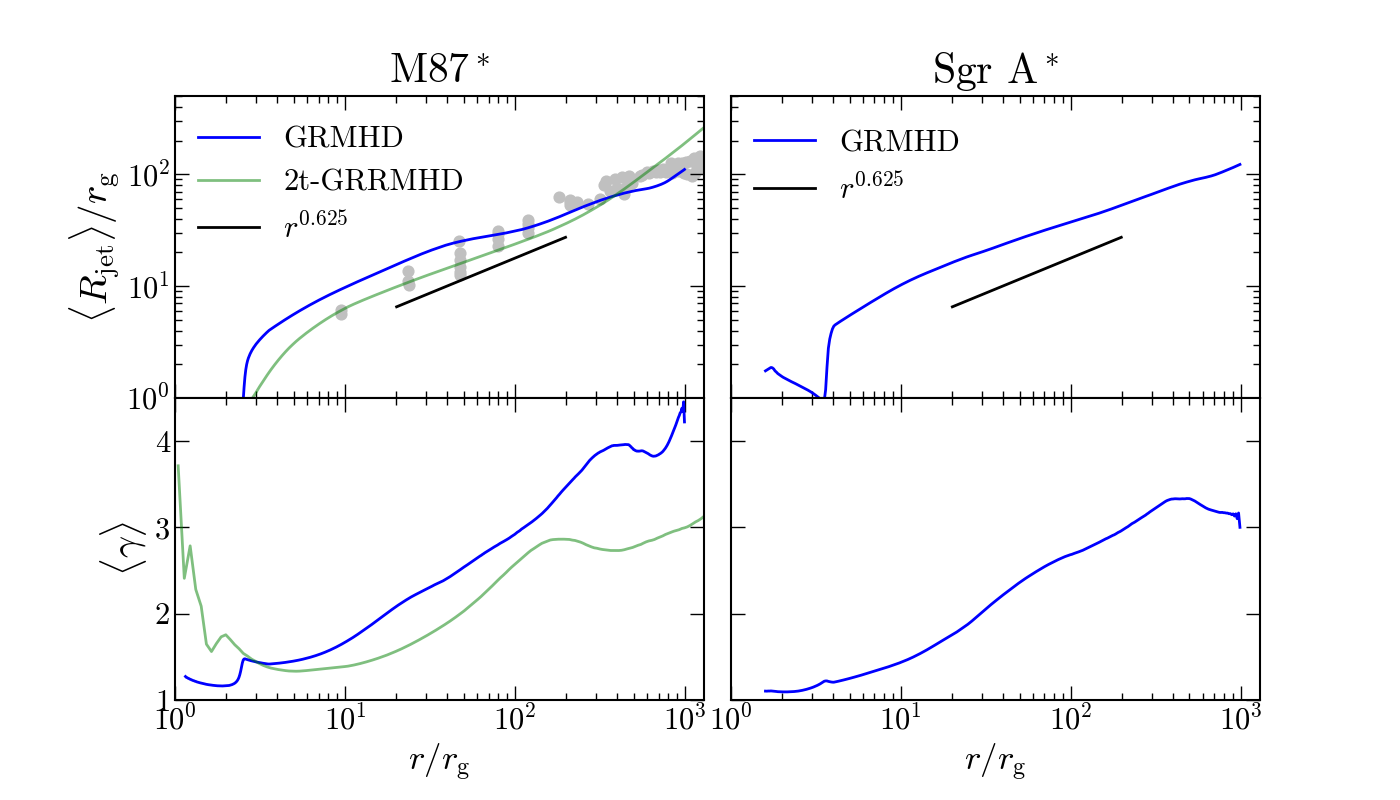}
	\caption{We show the jet radius $R_{\rm jet}$ and the jet Lorentz factor $\gamma$ from the \MBH{} GRMHD and 2t-GRRMHD models, and the \sgra{} GRMHD model. The gray circles indicates the deprojected jet radius of the M87 jet assuming a BH mass of $6.2\times10^9M_{\odot}$ and a source inclination of $14^{\circ}$~\citep{Nakamura_2018}. The data points are a compilation of various papers~\citep{Doeleman2012,asadanak2012,Hada2013,nak2013,Akiyama:2015,Hada2016}.}
    \label{fig:jet_profiles}
\end{figure}

\subsection{Jet Properties}
Here, we calculate the radial profiles of the jet half width $R_{\rm jet}$ and Lorentz factor $\gamma$:
\begin{equation}
    R_{\rm jet} (r,t) =  \sqrt\frac{\int^{2\pi}_{0}\int^{\pi}_{0}\, \sqrt{-g/g_{rr}} (\mu>2) \, d\theta \, d\varphi}{2\pi}\,, 
    \label{eq:Rjet}
\end{equation}
\begin{equation}
    \gamma (r,t) = \frac{\int^{2\pi}_{0}\int^{\pi}_{0}\, (\mu>2) \, \alpha u^t \, \sqrt{-g} \, d\theta \, d\varphi}{\int^{2\pi}_{0}\int^{\pi}_{0}\, (\mu>2) \, \sqrt{-g} \, d\theta \, d\varphi},
\end{equation}
where $\mu=-T^r_t/(\rho u^r)$ is the specific radial energy density and $\alpha=1/\sqrt{-g^{tt}}$. We calculate the average jet radius $R_{\rm jet}$ by integrating over the total surface area covered by the two jets over a shell at a particular radius, and assume that the total surface area is approximately $2\pi R_{\rm jet}^2$. We define the jet boundary as $\mu>2$, i.e., the region over which the jet still remains highly energized. This definition of the jet boundary is quite similar to the condition $\sigma=1$, which is widely used in the literature (e.g.,~\cite{Narayan2022}). Since $\mu=\gamma(\sigma+E_{\rm th})$, our condition $\mu>2$ also incorporates regions where the jet might not be magnetically dominated but is relativistically hot or fast. Since we restrict our jet profiles to within $r\lesssim 10^3\,r_{\rm g}$, the jet radius is primarily determined by the jet magnetization. \endnote{Specific enthalpy includes the rest-mass energy contribution in our definition from Section~\ref{sec:two-temp}.}

Figure~\ref{fig:jet_profiles} shows the jet radius $R_{\rm jet}$ and Lorentz factor $\gamma$ as a function of the radial distance from the BH for the \MBH{} GRMHD and 2t-GRRMHD models as well as the \sgra{} GRMHD model. The \MBH{} jet radius from our models matches the observed jet width from M87 (gray circles) quite well, with the radial profile roughly proportional to $r^{0.625}$, which is the fitted powerlaw for the M87 jet~\citep{Nakamura_2018}, though the index value has also been reported to be slightly smaller in some works (0.57;~\cite{asadanak2012,Nokhrina:2019}). The power law index of 0.625 is larger than that found using the $\sigma=1$ condition from~\citet{Narayan2022}, where the authors found a power law index of 0.428 for their MAD spin $a=0.9$ GRMHD model. It is possible that we find larger jet radii as we incorporate a part of the hot jet sheath region within our definition of $R_{\rm jet}$ (as suggested by Figure~7 in~\cite{chatterjee2019}). For the \sgra{} model, we also find a similar $R_{\rm jet}$ profile. While there are no detections of an extended jet in \sgra{} (e.g.,~\cite{Issaoun_2019}), semi-analytical and GRMHD models of \sgra{} largely favor a jet component from a spinning BH (e.g.,~\cite{Markoff:07,EHT_SgrA_2022_PaperV}).\vspace{-3pt}

We also show the Lorentz factor $\gamma$ in Figure~\ref{fig:jet_profiles}. Mostly, the jets accelerate to $\gamma\approx$ 3--4 by $10^3\,r_{\rm g}$ in all of our GRMHD models. It is difficult to compare our $\gamma$ profiles with values inferred from observations of the M87 jet (e.g.,~\cite{mertens2016}) since our $\gamma$ values are biased toward the jet spine while the observations generally capture the velocities of localized features in the sub-relativistic jet sheath/disk wind, especially at small distances from the BH. Indeed, both simulations and observations show that the jet Lorentz factor varies greatly as a function of jet radius (e.g., see~\cite{chatterjee2019}). We speculate that a better approach might be to calculate emissivity-weighted Lorentz factors in order to compare to the measured $\gamma$ from M87. Since our focus is on the comparison between GRMHD simulations, we leave direct comparisons to observational data to future work. 

\subsection{Axisymmetrized Profiles}

\begin{figure}[H]
    \includegraphics[width=\textwidth]{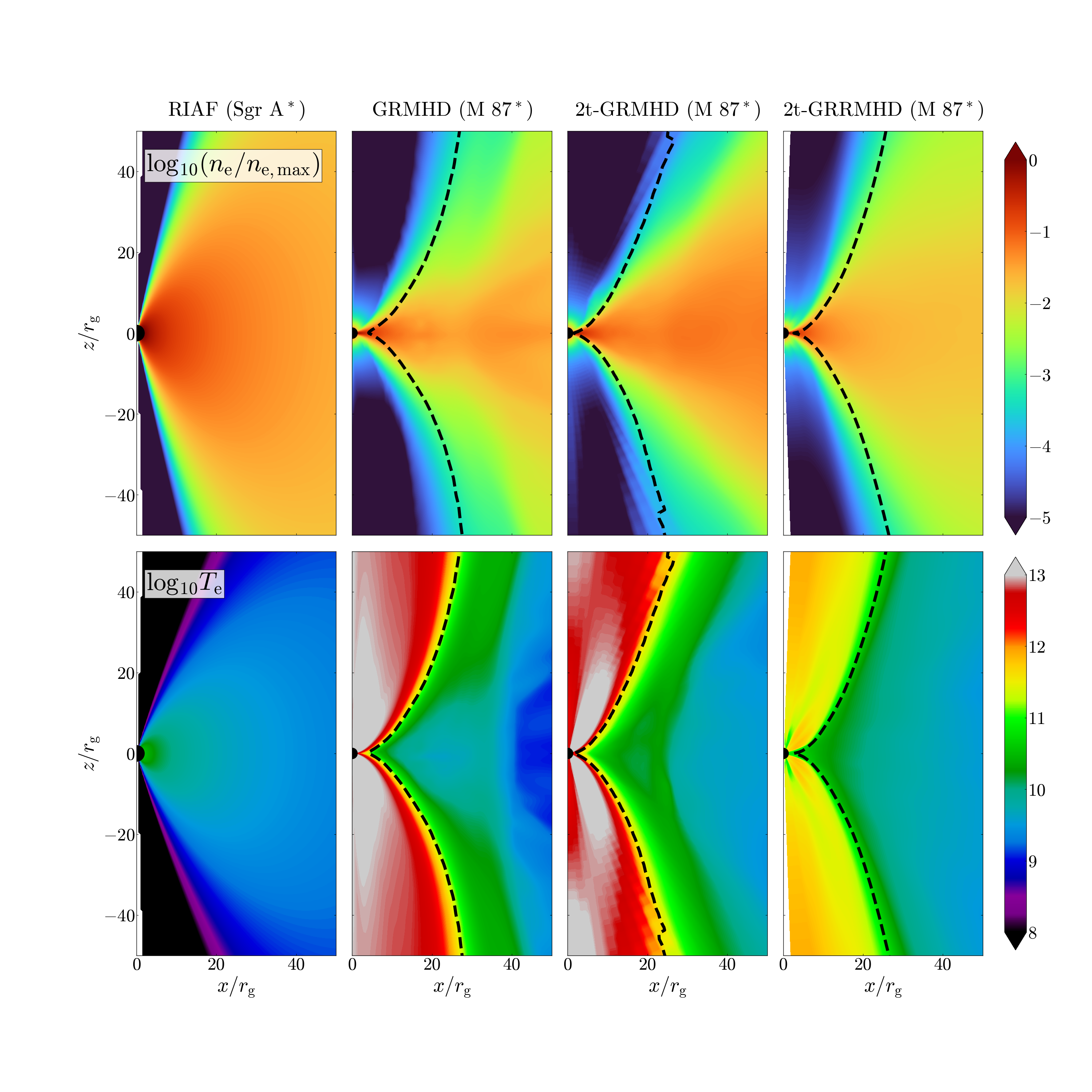}
    \vspace{-30pt}
	\caption{We show t- and $\varphi$-averaged data: electron number density $n_{\rm e}$ (top row) and temperature $T_{\rm e}$ (bottom row). We also denote the jet boundary with $\sigma=1$ (black lines). The time-averaging is performed over the $5000\,r_{\rm g}/c$ for each model. RIAF plots are for Sgr A$^*$ while the rest are for M87. The \sgra{} GRMHD model produces similar plots of $n_{\rm e}$ and $T_{\rm e}$ as the \MBH{} model, and, hence, we do not show it here.}
    \label{fig:2D_profiles}
\end{figure}

In the previous sections, we found that the largest differences between the GRMHD models occur in electron temperature distribution. Figure~\ref{fig:2D_profiles} shows the time and azimuthally averaged 2D vertical plots of gas density $n_{\rm e}$ and electron temperature $T_{\rm e}$. We show the normalized $n_{\rm e}$ so as to capture the relative change in the disk/wind density distribution, which would provide us information about the disk structure. The large difference in disk scale height is immediately apparent between the RIAF and the MAD GRMHD models (also see Figure~\ref{fig:disk_profiles_2}). The presence of a prominent wide jet component in MADs squeezes the inner disk and pushes against the accretion flow, a feature which is not captured in the constant $h/r$ RIAF model. However, the RIAF model roughly reproduces the density profile of the disk midplane region, suggesting that the RIAF model could represent non/weakly jetted, quasi-spherical accretion flows quite well. For sources such as \MBH{}, where we see a prominent jet component, the density gradient in the vertical direction is expected to be steeper as strong magnetic stresses power winds carry gas away from the disk (e.g.,~\cite{Chatterjee:2022}).

Overall, the disk/wind density distribution among the GRMHD models look similar with small differences in the lateral extension of the wind region and the steepness of the vertical gradient in density. For example, if we compare the 2t-GRRMHD model with the other two simulations, the density in the wind region is larger in the radiative model. The reason for the shallow vertical density profile in the 2t-GRRMHD model is unclear since the weakly magnetized thick disk simulations tell us that radiative cooling would lead to the loss of gas pressure in the disk and would result in the disk collapsing to a relatively dense structure in the midplane (e.g.,~\cite{fm09,Yoon:2020}). However, in the presence of strong poloidal magnetic fields, i.e., in the MAD state, the plasma-$\beta$ decreases to $\beta \approx$ 0.2--1 in the disk midplane (see Figure~\ref{fig:disk_profiles_1}, third row, left panel), and can reach even lower values in the upper layers of the accretion flow. The high magnetic pressure could help support the disk against collapse while sufficiently strong magnetic stresses could power disk winds. Such behavior is also seen in recent GRMHD simulations of near-Eddington, geometrically thin, strongly magnetized disks, where the inner disk (or corona) has a larger $h/r$ than the outer disk due to magnetic pressure support~\citep{Sadowski:2016_thin, Liska:2022, MishraB:2022}. To verify how radiative cooling affects the inner disk/wind structure in highly sub-Eddington accretion flows such as \MBH{} and \sgra{}, we require longer 2t-GRRMHD simulations such that the disk is in an inflow--outflow equilibrium out to a radius of at least $50\,r_{\rm g}$.

\vspace{-30pt}

The 2D temperature plot of the RIAF model also looks vastly different in the inner disk ($r\lesssim20\,r_{\rm g}$) when compared to the GRMHD and 2t-GRMHD simulations, but is similar to the temperature distribution in the 2t-GRRMHD disk midplane (also seen in the $T_{\rm e}$ plot of Figure~\ref{fig:disk_profiles_1}). The RIAF model does not capture gas heating in the jet sheath region (the region just outside of the jet boundary indicated by the $\sigma=1$ dashed line) and, therefore, $T_{\rm e}$ drops as we move away from the midplane toward the poles. In the GRMHD models, the jet sheath is as hot, if not hotter, than the inner accretion flow as temperatures reach $T_{\rm e}>10^{11}$~K. For the GRMHD simulation, the electron temperature is provided as a fraction of the fluid temperature, where the fraction depends on how magnetized the gas is in the region, as per the $R-\beta$ prescription from Equation~(\ref{eq:Rbeta}). For the \MBH{} model, we chose a $R_{\rm high}$ value of 160 to have a jet-dominated sub-mm image. This choice of $R_{\rm high}$ suppresses the electron temperature in the disk, focusing higher temperatures in the jet sheath. Comparing the GRMHD model with the 2t-GRMHD model, the jet sheath region exhibits very similar $T_{\rm e}$ values, but the disk midplane is hotter by a factor of a few in the 2t-GRMHD model. We note that this difference in $T_{\rm e}$ in the midplane is more noticeable in the 2D plot rather than in the disk-averaged $T_{\rm e}$ profile shown in Figure~\ref{fig:disk_profiles_1} as the upper layers of the disk become substantially hotter in the GRMHD model.

For the radiative 2t-GRRMHD model, the inner regions of the disk are cooler as electrons heated by magnetic reconnection quickly cool via synchrotron and Compton losses. From Figure~\ref{fig:disk_profiles_1}, the drop in $T_{\rm e}$ for the 2t-GRRMHD model is shown to be as large as an order of magnitude when compared to the (2t-) GRMHD models. Another interesting feature is that the hot region ($T_{\rm e}>10^{11}$ K) in the jet sheath is much narrower in the 2t-GRRMHD model, which could have a significant bearing on the ray-traced image, possibly producing a thinner jet sheath. Finally, the difference in $T_{\rm e}$ in the jet body between the GRMHD models is due to the different density/internal energy floor setups used by the corresponding codes. Since the gas in the jet sheath and the jet body undergo mixing due to boundary instabilities (e.g.,~\cite{chatterjee2019, Wong:2021}), it is possible that the choice of floors could affect the overall electron temperature in the jet sheath. Such a study is outside the scope of our paper and is left to future work.

\vspace{-30pt}
\begin{figure}[H]
    \includegraphics[width=\textwidth]{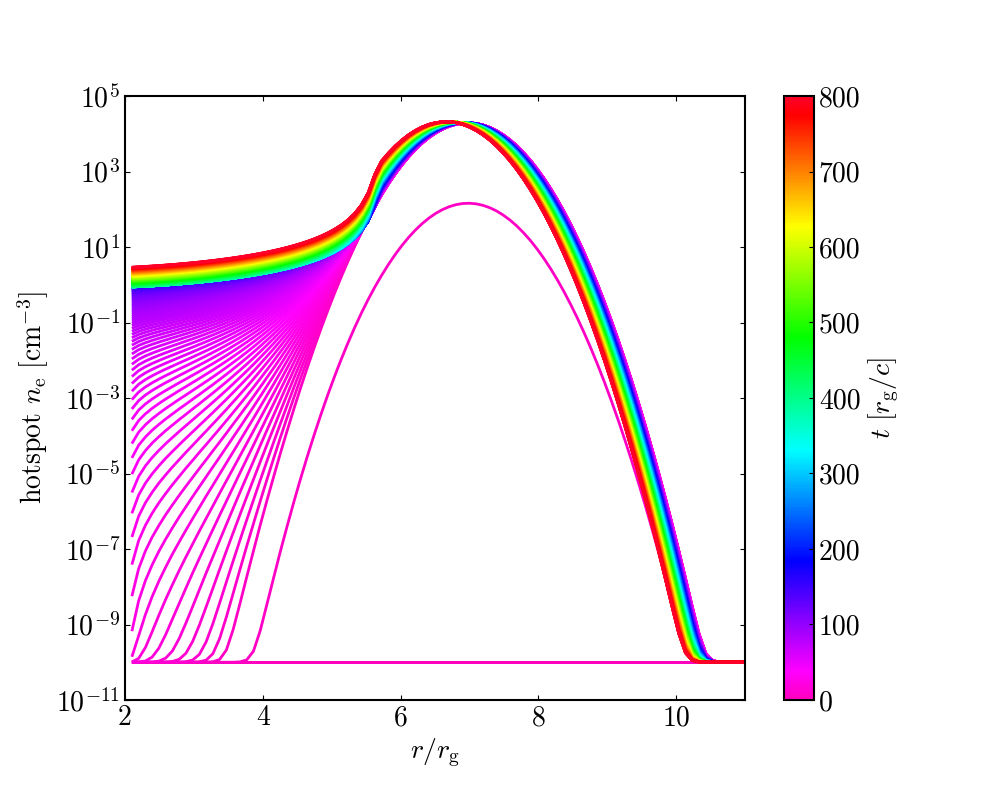}
	\caption{We show the $\varphi-$averaged hotspot electron number density as a function of radius and time. The hotspot falls into the BH and becomes sheared over time.}
    \label{fig:hotspot}
\end{figure}

\subsection{Orbiting Hotspot in a RIAF Model}

High-energy flares are commonly observed in AGNs, with GeV and MeV flares seen in \MBH{} (e.g.,~\cite{Aharonian:2006, Acciari:2010}) and quasi-daily nIR and X-ray flares in \sgra{} (e.g.,~\cite{Baganoff:01,Eckart:2006:flare_activity,Hornstein_2007,Nowak_2012,Neilsen_2013,Witzel_2018,Do_2019,Haggard_2019}). A number of attempts have been performed to explain the origin of flaring, such as magnetic reconnection in turbulent gas flows in the disk and the jet~\citep{Dodds-Eden_2010, Dibi_2014, chatterjee2019,Nathanail_2020,Chatterjee:2021} and magnetic flux eruptions~\citep{Dexter:2020:NIR_MAD,Porth:2020:NIR_MAD,Scepi:2022,Ripperda2022}. For \sgra{}, the semi-analytical models found that we require high-energy electrons, assumed to be accelerated via an ad hoc process such as a large magnetic reconnection event or shocks, to describe the large flaring events~\citep{Markoff_2001_sgra, Dibi_2016, Gutierrez_2020}. The near-infrared observations from the GRAVITY telescope provided further evidence for orbiting hotspot features in the accretion flow~\citep{Gravity:20:orbit} that may be linked to acceleration events. It has also been recently shown that orbiting hotspots can be used to model double-peaked X-ray flares~\citep{Haggard_2019, Ball_2021} and prominent Stokes Q-U loops in the sub-mm emission of \sgra{}~\citep{Wielgus:2022}. These results provide us with considerable motivation to test the capability of the ngEHT to detect hotspot formation in accretion flows around black holes.

Instead of isolating a particular magnetic flux eruption event in our simulations, we added a shearing hotspot to the RIAF solution as detailed in Section~\ref{sec:RIAF}. Figure~\ref{fig:hotspot} shows the temporal evolution of the azimuthally averaged electron number density of the hotspot. We begin with a Gaussian distribution of gas that undergoes shearing as the gas falls in closer to the BH. The overall density normalization is much lower than in the RIAF disk since the optically thin hotspot gas produces a large enough non-thermal synchrotron emissivity. The hotspot is evolved over $800\,r_{\rm g}/c$, but the gas distribution comes to a near-steady-state profile within the first $200\,r_{\rm g}/c$, which is roughly one hour for \sgra{}. The shearing of the hotspot gas has a significant impact on the evolution of the 230 GHz image~\citep{Tiede2020, Roelofs_ngEHT}. From Figure~\ref{fig:disk_profiles_2} (right column), we see that the radial velocity matches the disk-averaged gas velocity from the GRMHD model, showing nearly free-fall speeds, while the azimuthal velocity becomes highly sub-Keplerian. The velocity profiles show that our hotspot model should be able to reproduce the expected hotspot motion from the GRMHD models, and is ideal for investigating multiwavelength flare lightcurves. \citet{Emami:2022_hotspot}, a companion paper, goes into further details about how current dynamical reconstruction techniques can be used to trace out the motion and morphology of the shearing hotspot in the context of ngEHT observations. These hotspot models and reconstruction methods would be integral in deciphering the more complex gas dynamics of magnetic flux eruption events in MADs, which have been shown to produce a significant variation in the image structure of \MBH{} at 230 GHz (e.g.,~\cite{Gelles:2022}).

\section{Conclusions}
\label{sec:conclusions}

In this work, for the first time, we compare a series of numerical solutions with increasing complexity that were specifically constructed to understand the accretion flow around the supermassive black holes \MBH{} and \sgra{}. We include a time-independent radiatively inefficient accretion flow (RIAF) model as well as fully 3D GRMHD simulations of accreting black holes, incorporating the effects of electron heating and cooling losses via two-temperature and radiation physics. In addition, each of our simulations are run with different GRMHD codes, which is similar to the approach of another community-wide code comparison effort~\citep{Porth:19}. We found that the simulations exhibit remarkably similar properties given that the simulations incorporate varying levels of complexity in electron physics. The notable exception is the electron temperature, where radiative cooling decreases the temperature by a factor of $\lesssim 10$ within the inner 10 gravitational radii, the region that produces the bulk of the 230 GHz emission in \MBH{}, one of the two primary targets of the EHT and the ngEHT (the other being \sgra{}). The main goal of this work is to understand the variation in the underlying accretion flow and jet properties in our models since synthetic ray-traced images constructed from these models are used as ``truth'' images of \MBH{} and \sgra{} for the ngEHT Analysis Challenges~\citep{Roelofs_ngEHT}. The ngEHT Analysis Challenges are an effort to determine how much information about the accretion flow and jet dynamics we can glean from the proposed ngEHT reference array, and what modifications in the image reconstruction tools are necessarily required to decode future ngEHT observational~data. 

Our paper deals with numerical models designed to investigate hotspot evolution, turbulent inspiralling gas flows, and extended powerful jets, targeting \MBH{} and \sgra{}. We restricted our model set to the community-standard setup: a rotating, geometrically thick, optically thin torus of magnetized gas around a spinning black hole, which is the fiducial model choice of the EHT~\citep{EHT_M87_2019_PaperV, EHT_M87_2019_PaperVIII, EHT_SgrA_2022_PaperV}. This model choice leaves out the exploration of multiple new setups of black hole accretion, such as quasi-spherical wind-fed inflows (e.g.,~\cite{Ressler:2020:sgra_MAD, Lalakos:2022}), strongly wind-fed accretion (e.g.,~\cite{Cruz-Osorio:2017,Kaaz:2022}), geometrically thin accretion disks (e.g.,~\cite{Avara:2016, Liska:2022, MishraB:2022}), puffy radiation-dominated super-Eddington disks (e.g.,~\cite{Sadowski:2016, Curd:2019}), and misaligned accretion disks (e.g.,~\cite{fragile07, liska_tilt_2018, White_2019, Chatterjee_2020}). Apart from varying the accretion mode, the high resolution of the images from EHT and ngEHT could potentially help distinguish between different space--time metrics~\citep{EHT_Sgra_paper_VI}. To date, only a limited number of non-Kerr GRMHD simulations have only been performed (e.g.,~\cite{Mizuno:2018, Olivares:2020, Nampalliwar:2022PhRvD.106f3009N}). The future of numerical studies is bright, given their rising popularity in the astrophysics community and the increase in computational resources. The breadth of the current investigations in accretion physics would result in a plethora of variable structures that should be thoroughly studied keeping the observational capabilities of the ngEHT in mind.

\vspace{6pt}

\authorcontributions{Conceptualization, Koushik Chatterjee; Methodology, Koushik Chatterjee, Paul Tiede, Yosuke Mizuno and Freek Roelofs; Software, Koushik Chatterjee, Andrew A. Chael, Paul Tiede and Yosuke Mizuno; Formal analysis, Koushik Chatterjee and Paul Tiede; Investigation, Koushik Chatterjee, Yosuke Mizuno, Razieh Emami, Christian Fromm, Angelo Ricarte, Lindy Blackburn, Freek Roelofs, Philipp Arras, Antonio Fuentes, Jakob Knollmüller, Nikita Kosogorov, Greg Lindahl, Hendrik Müller, Nimesh Patel, Alexander Raymond, Efthalia Traianou and Justin Vega; Writing – original draft, Koushik Chatterjee and Paul Tiede; Writing – review \& editing, Koushik Chatterjee, Andrew A. Chael and Angelo Ricarte; Visualization, Koushik Chatterjee; Supervision, Michael D. Johnson and Sheperd Doeleman.
}

\acknowledgments{We thank the anonymous referees for their suggestions that greatly improved the text. This research was enabled by support provided by grant no. NSF PHY-1125915 along with INCITE and ASCR Leadership Computing Challenge (ALCC) programs under the award PHY129, using resources from the Oak Ridge Leadership Computing Facility, Summit, which is a US Department of Energy office of Science User Facility supported under contract DE-AC05- 00OR22725, as well as Calcul Quebec and Compute Canada under award xsp-772-ab (PI: D. Haggard).}

\funding{We thank the National Science Foundation (AST-1716536, AST-1935980, AST-1935980, and AST-2034306) and the Gordon and Betty Moore Foundation (GBMF-10423) for financially supporting this work. This work was supported in part by the Black Hole Initiative, which is funded by grants from the John Templeton Foundation (JTF-61497) and the Gordon and Betty Moore Foundation (GBMF-8273) to Harvard University. K.C. is also supported in part by the Black Hole PIRE program (NSF grant OISE-1743747). R.E. acknowledges the support by the Institute for Theory and Computation at the Center for Astrophysics as well as grant numbers 21-atp21-0077, NSF AST-1816420, and HST-GO-16173.001-A for very generous supports. H.M. received financial support for this research from the International Max Planck Research School (IMPRS) for Astronomy and Astrophysics at the Universities of Bonn and Cologne. This research is supported by the DFG research grant ``Jet physics on horizon scales and beyond'' (Grant No.  FR 4069/2-1), the ERC synergy grant ``BlackHoleCam: Imaging the Event Horizon of Black Holes'' (Grant No. 610058), and ERC advanced grant ``JETSET: Launching, propagation and emission of relativistic jets from binary mergers and across mass scales'' (Grant No. 884631). J.K. acknowledges funding by the Deutsche Forschungsgemeinschaft (DFG, German Research Foundation) under Germany´s Excellence Strategy---EXC 2094---390783311. Y.M. is supported by the National Natural Science Foundation of China (No. 12273022) and the Shanghai pilot program of international scientists for basic research (No. 22JC1410600).}

\dataavailability{The data shown in this study has been generated using the following codes: \bhac{}~\citep{porth17}, \bhoss{}~\citep{Younsi:19_polarizedbhoss}, \hamr{} \citep[][]{liska_hamr2020}, \ipole{}~\citep{Moscibrodzka:2018}, \koral{}~\citep{Sadowski:13_koral}, with access provided upon request to the corresponding authors.}

\conflictsofinterest{The authors declare no conflict of interest. The funders had no role in the design of the studyl in the collection, analyses, or interpretation of data; in the writing of the manuscript, or in the decision to publish the results.}

\begin{adjustwidth}{-\extralength}{0cm}
\printendnotes[custom]

\reftitle{References}

\PublishersNote{}
\end{adjustwidth}
\end{document}